\documentclass[1p,12pt]{article}
\newcommand{\ZZL}{\langle\!\langle}
\newcommand{\ZZR}{\rangle\!\rangle}

\usepackage{color}
\usepackage{amssymb,amsmath,exscale}
\def\zsumU{\sum _{n=1}^{\ZIN}}

\def\zsumNoz{\sum _{n\neq  0} }

\newcommand{\ZEP}{\epsilon}

\newcommand{\zben}{\beta_n}
\newcommand{\zg}{\gamma}
\newcommand{\zn}{Z_n}

\newcommand{\intT}{\int_0^T}
\newcommand{\intt}{\int_0^t}

\newcommand{\intr}{\int_0^r}

\newtheorem{Theorem}{Theorem}
\newtheorem{Corollary}[Theorem]{Corollary}
\newtheorem{Lemma}[Theorem]{Lemma}

\newtheorem{Remark}[Theorem]{Remark}
\newcommand{\zdiaform}{\mbox{~~\zdia}}

\newcommand{\ZOM}{\omega}
\newcommand{\zaa}{\alpha}

\newcommand{\ZDE}{\delta}
\newcommand{\zt}{\tau}
\newcommand{\zdia}{~~\rule{1mm}{2mm}\par\medskip}
\newcommand{\zthe}{\theta}

\newcommand{\ZLA}{\label}
\newcommand{\ZIN}{\infty}
\newcommand{\zProof}{{\noindent\bf\underbar{Proof}.}\ }

\newcommand{\ZBI}{\bibitem}
\newcommand{\ZD}{\;\mbox{\rm d}}

\newcommand{\ZSI}{\sigma}

\author{
L. Pandolfi\thanks{Dipartimento di Scienze Matematiche, Politecnico di Torino, Corso Duca degli Abruzzi 24---10129 Torino, Italy (luciano.pandolfi@polito.it)}
}
\title{Traction, deformation and velocity of deformation in a viscoelastic string\thanks{
This papers fits into the research program of the GNAMPA-INDAM and has been written in the framework of the   ``Groupement de Recherche en Contr\^ole des EDP entre la France et l'Italie (CONEDP-CNRS)''. The  author is partially supported by Italian MURST.}}

\begin{document}
\maketitle
 
\begin{abstract}

In this paper we consider a viscoelastic string whose deformation is controlled at one end. We study the relations and the controllability of the couples traction/velocity and traction/deformation and we show that the first couple behaves very like as in the purely elastic case, while new phenomena appears when studying the couple of the traction and the deformation. Namely, while traction and velocity are independent (for large time), traction and deformation are related at each time but the relation is not so strict. In fact we prove that an arbitrary number of ``Fourier'' components of the traction and, independently, of the deformation can be assigned at any time.

\end{abstract}



\section{Introduction}

Let a viscoelastic string be in eqilibrium on the interval $ [0,\pi] $ of the $ x $--axis. When a vertical displacement   is applied to the boundary point $ x=0 $, while the boundary point $ x=\pi $ is kept fixed, 
the dynamic of the string (at rest for negative times) is described by
\begin{equation}
\label{eq:equaSECord1}
w_{tt}(x,t)=w_{xx}(x,t)+\intt M(t-s) w_{xx}(x,s)\ZD s 
\end{equation}
  with initial and boundary conditions
  \begin{equation}
  \label{eq:icb}
  \left\{ \begin{array}{ll}
  w(x,0)=0 ,  \quad
w_t(x,0)=0\\
  w(0,t)= f(t)\in L^2 _{\rm loc}(0,+\ZIN)\,, &w( \pi,t)=0\,.  
\end{array} \right.
\end{equation}
  Here $ w(x,t) $ denotes the vertical displacement and $ w_t(x,t) $ the vertical velocity of the point in position $ x $ at time $ t $. The  vertical components of the stress     at position $ x $ and time $ t $ is
\begin{equation}\ZLA{eq:DEFIstress}
\ZSI(x,t)=  w_x(x,t)+\intt M(t-s) w_x(x,s)\ZD s
\end{equation}
(the usual minus sign in front of the right hand side has no interest for the following and we drop it).

We shall assume that the real kernel $ M(t) $ is  of class $ H^2_{\rm loc}(0,+\ZIN) $  and we note that in general there will be a certain (positive) coefficient in front of the laplacian $ w _{xx} $, which has been taken equal to $ 1 $ for simplicity.
 
  In order to understand the problem that we are going to study in this paper, let us first consider the purely elastic case, i.e. the case $ M(t)\equiv 0 $. In this case Eq.~(\ref{eq:equaSECord1}) is the string equation and
   it is known that:
   \begin{itemize}
   \item for every $ T>0 $ we have: $ w(t)\in C(0,T;L^2(0,\pi))$ (and $ f\mapsto w $ is continuous from $ f\in L^2(0,T) $); $  w_t(t)\in C(0,T ;H^{-1}(0,\pi)) $  (and $ f\mapsto w_t $ is continuous from $ f\in L^2(0,T) $);
   \item     for every target $ \xi \in L^2(0,\pi) $, $ \eta \in H^{-1}(0,\pi) $ it is possible to find a control $ f\in L^2(0,T) $
 such that
 \[ 
 w(T)=\xi\,,\qquad w_t(T)=\eta
  \]
  provided that $ T\geq 2\pi $.
   \end{itemize}
 Note that we suppress the dependence on the state variable $ x $, unless needed for clarity, so that $ w(t) $ denotes $ w(x,t) $. Moreover, $ w $ does depend on $ f $ but this is not explicitly indicated.
 
  Consequently, we have also  $ w_x(t)\in C(0,T;H^{-1}(0,\pi)) $ for every $ T>0 $ and the fact that $ \xi $ and $ \eta  $ are arbitrary shows that also the stress/ve\-locity pair $ (w_x(T),w_t(T))\in H^{-1}(0,\pi)\times H^{-1}(0,\pi) $ can be arbitrarily assigned under the action of the boundary control, when $ T\geq 2\pi $. This observation can be interpreted both as a controllability property of the wave equation and as the fact that stress and velocity at a certain time $ T $ (large enough) are independent   (which seems to us the most enlightening interpretation). Instead the pair $ (w(T),w_x(T)) $ cannot be controlled: in fact the displacement identifies the stress (Hooke Law).
  
  Now we consider the viscoelastic string. We proved in~\cite{LoretiPANDOLFIsforza} that the pair $ (w(t),w_t(t)) $ has the same property as in the purely elastic case: it belongs to $ C(0,+\ZIN;L^2(0,\pi)\times H^{-1}(0,\pi)) $ and its value at a certain time $ T $ can be assigned at will in  $L^2(0,\pi)\times H^{-1}(0,\pi)$ provided that $ T $ is large enough ($ T\geq 2\pi $). In this paper, using   results in~\cite{AvdoninPANDOLFI1,AvdoninPANDOLFI2}, we first   prove that, in spite of the memory term, also the pair $(\ZSI(x,T),w_t(x,T))   $ can be arbitrarily assigned for $ T\geq 2\pi $. 
  
  \begin{Theorem}\ZLA{teo:controllSTRESS/velocita}
Let $ \ZSI(t) $ be defined in~(\ref{eq:DEFIstress}). Then we have:
\begin{enumerate}
\item The function $ f\mapsto  w_t( t) $ is linear and continuous from $ L^2(0,T) $ to $ C(0,T;H^{-1}(0,\pi) )$ for every $ T>0 $;
\item The function $ f\mapsto  \ZSI( t) $ is linear and continuous from $ L^2(0,T) $ \\ to $H^{-1}(0,T;L^2(0,\pi))\cap C(0,T;H^{-1}(0,\pi)) $ for every $ T>0 $;
\item
Let $ \xi\in H^{-1}(0,\pi) $, $ \eta\in H^{-1}(0,\pi) $ and let $ T\geq 2\pi $. Then there exists $ f\in L^2(0,T) $ such that
\[ 
\ZSI(T)=\xi\,,\qquad w_t(T)=\eta\,.
 \]
 \end{enumerate}
\end{Theorem}

Statement  1) is proved in~\cite{LoretiPANDOLFIsforza}   while we shall see below, see Lemma~\ref{eq:lemmaCHEproVA1-2Teo1},  that statement 2) follows from~\cite{AvdoninPANDOLFI1}.
So, we are mostly interested in the last statement, which can be   interpreted as controllability or independence of velocity and traction if the elapsed time is large enough.

Finally, in section~\ref{sect:stress-deformation} we shall study the pair  deformation/stress and we shall see that these functions are not independent, i.e. there is no controllability of that pair, and that a kind of ``Hooke Law'' holds asymptotically,
 for  short wavelength  components. But, we shall also see that, unlike the purely elastic case, the \emph{ long wavelength components    are  independent.} The precise statement is in section~\ref{sect:stress-deformation}.

\subsection{Comments on previous references}

Controllability properties of viscoelastic materials have been studied by several authors in past years, using different methods. See for example~\cite{BarbuIannelli,ZANGPRIMOLAVORO,Kim2,Leugering1984,Leugering1992,LoretiSforza,PandAMO,PandSISSA}. A constructive approach to the steering control (in the case of the heat equation with memory), based on moment methods, has been introduced in~\cite{PandIEOT} and then developped in subsequent papers~\cite{AvdoninPANDOLFI1,AvdoninPANDOLFI2,LoretiPANDOLFIsforza,PandDCDS1,PandDCDS2}. 
The key idea of these papers has been applied to a different class of problems in~\cite{ABP10}.

The papers~\cite{AvdoninPANDOLFI1,AvdoninPANDOLFI2} shows an interpretation of controllability of pairs of variables   as independence of that controlled variables. This approach we push further in this paper. We relay on the moment methods techniques introduced in the papers just cited, and in particular we shall use some results proved in~\cite{AvdoninPANDOLFI1,AvdoninPANDOLFI2,PandDCDS1}.

When studying distributed systems with memory, we might get the feeling that they are ``perturbations'' of heiter heat or wave equations, and behave much in the same way. This conjecture is disproved both from the results in Section~\ref{sect:stress-deformation} and the negative results in~\cite{GuerreroIMAN,HalanayPandolfi,IvanovPANDOLFI}.

\section{Preliminaries}

The following computations make sense for smooth boundary inputs $ f  $ and are then extended by continuity to $ f\in L^2(0,T) $.

Let
\[ 
 N (t)=1+\intt M(s)\ZD s
 \]
 (note that $ 1 $ is the coefficient of the  laplacian in~(\ref{eq:equaSECord1})). Then, integrating both the sides we can write~(\ref{eq:equaSECord1}) in the form
 \begin{equation}
\ZLA{eq:FORMAprimORDINE}
w_t=\intt   N (t-\tau) w_{xx}(\tau)\ZD \tau\,,\\
%w(x,0)=0\,,\qquad w(0,t)=f(t)\,,\quad w(\pi,t)=0\,.
%\end{array}
\end{equation}
We introduce
\begin{equation}
\ZLA{eq:perILflussoCALORE}
q(t)=q(x,t)=\intt \ZSI(x,\tau)\ZD \tau=\intt   N (t-\tau) w_x(x,\tau)\ZD \tau\,.
\end{equation}
This shows a relation with the first order systems studied in~\cite{AvdoninPANDOLFI1,AvdoninPANDOLFI2}.

For technical reasons, in the previous papers it proved convenient to replace $ w(x,t) $ with
 $
\zthe(x,t)=e^{2\zaa t}   w(x,t)
$
which solves the seemingly more involved equation, where
\[ 
N_{\zaa}(t)=e^{2\zaa t}N(t)\,:
 \]
 \begin{equation}
 \ZLA{eq:FORMAprimORDINENprimoNULLO}
 \begin{array}{l}
\displaystyle  \zthe_t= 2\zaa \zthe(t)+\intt    N _\zaa(t-\tau) \zthe_{xx}(\tau)\ZD \tau\,,\\[2pt]
\displaystyle \zthe(0)=0\,,\qquad \zthe(0,t)=e^{2\zaa t} f(t)\,,\quad \zthe(\pi,t)=0\,,\\[2pt]
\displaystyle \tilde q(x,t)=e^{2\zaa t}q(x,t)=-\intt  N_\zaa (t-\tau)\zthe_x(x,\tau)  \ZD \tau\,.
\end{array}
\end{equation}
This transformation has been introduced in~\cite{PandIEOT} and it turns out that it is convenient to choose
 \begin{equation}\ZLA{eq:lacondizioneNprimoZERO}
 \zaa=-\frac{1}{2}N'(0) \quad \mbox{so to have} \quad N_\zaa'(0)=0\,.
  \end{equation}

Now we can see that statement 2) of Theorem~\ref{teo:controllSTRESS/velocita} holds. In fact,  from~\cite{AvdoninPANDOLFI1},   $ \zthe\in C(0,T;L^2(0,\pi)) $ and $ q\in C(0,T;L^2(0,\pi)) $.    Hence we have
  \begin{Lemma}\ZLA{eq:lemmaCHEproVA1-2Teo1}
For every $ T>0 $ we have $ \ZSI\in C(0,T;H^{-1}(0,\pi)) \cap H^{-1}(0,T;L^2(0,\pi))$ and depends continuously on $ f\in L^2(0,T) $.
\end{Lemma}
In particular, this lemma shows that the stress, as an element of   $ H^{-1}(0,\pi) $, can be computed at each time $ t $.

The multiplicative transformation is innocuous since, with
$
M_\zaa(t)=e^{2\zaa t}M(t) 
$,
  we have
\begin{equation}
\ZLA{eq:defiSTRESSvelocityCONalpha}
\left\{\begin{array}{lll}
\displaystyle \ZSI(x,t)&=&e^{-2\zaa t}\left [ \zthe_x(x,t)+\intt M_\zaa(t-s) \zthe_x(x,s)\ZD s\right ]\,,\\
\displaystyle w_t(x,t)&=&e^{-2\zaa t}\left [ \zthe_t(x,t)-2\zaa \zthe(x,t)\right ]\,.
\end{array}\right.
\end{equation}

Projecting the solutions of Eq.~(\ref{eq:FORMAprimORDINENprimoNULLO}) on the spaces generated by $  \sqrt{2/\pi}\sin nx $ in $ L^2(0,\pi) $ we find the following representation/definition for the solutions of
 Eq.~(\ref{eq:FORMAprimORDINENprimoNULLO}) (see~\cite{PandIEOT}):
\[
 \zthe(x,t)=
 \zsumU \sqrt{\frac{2}{\pi} } (\sin nx)\zthe_n(t)
\]
where $ \zthe_n(0)=0 $ and  $ \zthe_n(t) $ solves
\begin{eqnarray*}
&& \zthe'_n(t) 
=2\zaa \zthe_n(t)-n^2\intt N_{\zaa}(t-s)\zthe_n(s)\ZD s+n v(t)\,,
\\
&& v(t)=\intt N_\zaa(t-s) \left (  \sqrt{\frac{2}{\pi} } e^{2\zaa s}  f(s)  \right )\ZD s\,.
\end{eqnarray*}

Let us introduce the solutions $ z_n(t) $ of the problem
 \begin{equation}
\ZLA{eq:diZnPRIMORD}
  z_n'(t)=2\zaa  z_n(t) -n^2\intt N_\zaa(t-s)   z_n(s)\ZD s\,,\qquad z_n(0)=1\,.
\end{equation}
Then we have (we rename $ f(t) $ the function $ e^{2\zaa t}  f(t)$)
\begin{eqnarray}
\nonumber&& \zthe_n(t)=\intt   z_n(t-s) \left (n v(s)\right )\ZD s\\
\ZLA{eq:formULAprRTheta}&&=\intt\sqrt{\dfrac{2}{\pi}} f(t-r)\left [n\int _0^{r}N_{\zaa}(r-s)z_n(s)\ZD s \right ]\ZD r
 \end{eqnarray}
 
So, the quantities of our interest are: 
 \begin{equation}\ZLA{serieDIflussoVELOCITA}
 \begin{array}{l}
 \displaystyle (\pi/2)e^{2\zaa t }  w(x,t)\\[3pt]
  \displaystyle =\zsumU (\sin nx)\intt f(t-\nu)\left [ n\int_0^{\nu}N_{\zaa}(\nu-s)z_n(s)\ZD s\right ]\ZD\nu\,,
 \\[7mm]
 \displaystyle (\pi/2)e^{2\zaa t}  w_t(x,t)  \\[3pt]
 \displaystyle =  \zsumU (n\sin nx)\intt f(t-\nu)\left [ z_n(\nu) 
 +\int_0^\nu H(\nu-s) z_n(s)\ZD s\right ] \ZD\nu\,,\\[7mm]
 \displaystyle (\pi/2)e^{2\zaa t} \ZSI(x,t) \\[3pt]
 \displaystyle = \zsumU (n\cos nx)\intt f(t-\nu)\left [
 n\left (
 \int_0^\nu K(\nu-s) z_n(s)\ZD s
 \right )
 \right ]  
 \end{array}
 \end{equation}
where
\begin{equation}\ZLA{eq:defiHK}
 H(t)= N_\zaa'(t)-2\zaa N_\zaa (t)\,,\qquad
  K(t)=N_\zaa(t)+\intt N_\zaa(t-r)M_\zaa(r)\ZD r\,.
\end{equation}

Convergence of the previous series in the appropriate spaces, $ C (0,T;L^2(0,\pi)) $ for the first and $ C(0,T;H^{-1}(0,\pi)) $ is known, see~\cite{AvdoninPANDOLFI1,LoretiPANDOLFIsforza,PandIEOT,PandDCDS1}.

It is clear from these formulas that control problems are easily reduced to moment problems. So, before we proceed, we present some background information on moment problems and Riesz sequences.

 \section{Preliminaries: moment problems, Riesz bases and Riesz sequences}
 
 Let $ H $ be a Hilbert space and $\{ h_n\} $ a (fixed) sequence in $ H $. Let us consider the infinite set of equations
\begin{equation}\ZLA{eq:MomeProbleASRTA}
 \langle u,h_n\rangle=c_n
  \end{equation}
  wher $ \langle\cdot,\cdot\rangle $ is the inner product in $ H $ and $ \{c_n\} $ is a sequence of complex number. Under the heading ``moment problem'' is intended the problem to caracterize those sequences $ \{ h_n\} $ such that a solution $ u  $ of the equations~(\ref{eq:MomeProbleASRTA}) exists for every
  sequence $ \{c_n\} $ with suitable properties. The key result of interest here is as follows (see~\cite[p.~34]{AvdoninIVANOV}).

  \begin{Theorem}
  The moment problem~(\ref{eq:MomeProbleASRTA}) is solvable for every sequence $ \{c_n\}\in l^2 $
  and the solution $ u\in H  $ depends continuously on $ \{c_n\}\in l^2 $
  if and only if $ \{h_n\} $ is a Riesz sequence in $ H $.  
  
  The solution is unique if $ \{h_n\} $ is a Riesz basis of $ H $.
  \end{Theorem}
  So, in order to make clear the content of this theorem, we must explain what a Riesz sequence is and in order to use it we need tests which can be used to see whether a sequence is Riesz. These are taken from~\cite[Ch.~1 sect.~9]{Young}.

 A sequence $ \{h_n\} $ in a Hilbert space $ H $ is a Riesz basis when there exists a linear bounded and boundedly invertible transformation $ \cal T $ in $ H $ such that $ h_n={\cal T}\ZEP _n $, where $ \{\ZEP_n\} $ is an orthonormal basis   of $ H $.
 
 If $ \{ h_n\} $ is a Riesz basis in its closed span then it is called a Riesz sequence.
 
 An equivalent condition is as follows:
 \begin{Theorem}\ZLA{Teo:BARIcarattRIESZ}
 A sequence $ \{h_n\} $ in a Hilbert space $ H $ is a Riesz sequence  if and only if   there exist  \emph{positive} numbers $ m $ and $ M $ such that for every finite sequence $ \{c_n\} $ of scalars we have      
\begin{equation}\ZLA{eq:Appe:condiRIesz}
m\sum |c_n|^2\leq\left \|\sum c_n h_n\right \|_{H}^2\leq M|c_n|^2\,.
 \end{equation}
 If  furthermore the sequence $ \{h_n\}  $ is complete, then it is a Riesz basis, and conversely.
 \end{Theorem}
 
 Let $ \{h_n\} $ be a   sequence in $ H $. A Paley-Wiener theorem, adapted to Hilbert spaces and orthonormal bases, states that if $ \{e_n\} $ is an orthonormal basis of $ H $ and
 \[ 
 \sum \|h_n-e_n\|^2<1
  \]
  then $ \{h_n\} $ is a Riesz basis.  A corollary which will be used is as follows:
  \begin{Corollary}\ZLA{coro:estePaleyWIE}
  Let $ \{e_n\} $ be a Riesz sequence and let the sequence $ \{h_n\} $ satisfy
  \begin{equation}
\ZLA{eq:PaleyWien}
 \sum \|h_n-e_n\|^2<+\ZIN
    \end{equation}
  then there exists a number $ N $ such that $ \{ h_n\} _{n>N} $ is a Riesz sequence too. Consequently, if~(\ref{eq:PaleyWien}) holds then
\[%\begin{equation}
\ZLA{eq:PrimaCONDbari}
 \sum \zaa_n h_n
\]%\end{equation}
  converges in the norm of $ H $ if and only if $ \{\zaa_n\}\in l^2 $.
   \end{Corollary}
   We stress that the sequence $ \{e_n\} $ in Corollary~\ref{coro:estePaleyWIE} need not be an orthonormal basis.
   
Condition~(\ref{eq:PaleyWien}) does not imply that $ \{h_n\} $ is a Riesz sequence but 

\begin{Theorem}[Bari Theorem] If both the condition~(\ref{eq:PaleyWien}) and the condition~(\ref{eq:deFiZOMindep}) below hold then $ \{h_n\} $ is a Riesz sequence. 
\end{Theorem}
The additional condition~(\ref{eq:deFiZOMindep}) is called {\em $ \ZOM $-independence\/} and it is
\begin{equation}
\ZLA{eq:deFiZOMindep}
\sum \zaa_n h_n=0\ \implies\ \{\zaa_n\}=0\,.
\end{equation}
The convergence of the series is in $ H $ so that, as noted in Corollary~\ref{coro:estePaleyWIE}, the convergence of the series in~(\ref{eq:deFiZOMindep}) implies $ \{\zaa_n\}\in l^2 $.
 
 Finally, we state the following lemma. For completeness, we give a proof in Appendix~\ref{appendix:prooflemma:complettezzaH-1}.
 
\begin{Lemma}\ZLA{lemma:complettezzaH-1}
The sequence $ \{(\sqrt{2/\pi})n\sin nx\}_{n\geq 1}  $ is an orthonormal  basis in $ H^{-1}(0,\pi) $ while $ \{n\cos nx\}_{n\geq 1}  $ is a Riesz basis  in $ H^{-1}(0,\pi) $.
\end{Lemma}

 %%%%%%%%%%%%%%%%%
 \section{The   stress and the velocity}
 In this section we consider the pair  stress/velocity  and we prove Theorem~\ref{teo:controllSTRESS/velocita}. We proceed in several steps.

Lemma~\ref{lemma:complettezzaH-1} shows that
 every pair $ (\xi,\eta)\in H^{-1}(0,\pi)\times  H^{-1}(0,\pi)$
can be represented as
\[ 
\xi=\sum _{n-1}^{+\ZIN} \xi_n\left (n\sin nx\right )\,,\qquad
\eta=\sum _{n-1}^{+\ZIN} \eta_n\left (n\cos nx\right ) 
 \]
 where
 \[ 
 \{\xi_n\}\in l^2\,,\qquad \{\eta_n\}\in l^2\,,
  \]
  and conversely.
 
Hence, given $ T>0 $, the pair $ (\xi,\eta) $ in $ H^{-1}(0,\pi)  $ is reachable at time $ T $ by the pair $ \left (w_t(\cdot,T), \ZSI(\cdot,T)\right ) $ if   the
 following moment problem is solvable (see~(\ref{serieDIflussoVELOCITA}).   We ignore the inessential factor $(\pi/2) e^{2\zaa T}  $):
\begin{equation}
\ZLA{eq:MOMENTproblemDASTUDIARE}
\left\{\begin{array}{l}
\displaystyle \intT f(T-r)\left \{ z_n(r)+\intr H(r-s) z_n(s)\ZD s  \right \}  \ZD r=\xi_n\,,\\
\displaystyle  \intT 
f(T-r)\left \{ n\intr K(r-s)z_n(s)\ZD s\right \}\ZD r
=\eta_n\,.
\end{array}
\right.
\end{equation}
So, our goal is the proof  that this moment problem is solvable for arbitrary 
  sequences $ \{(\xi_n,\eta_n)\} $  in $ l^2\times l^2 $, i.e. arbitrary $ \{\xi_n+i\eta_n\} $  in $ l^2_{\mathbb C}  $, the $ l^2 $-space of complex valued sequences (and $ n $ is natural, $ n>0 $). 
Even more, we prove that the solution $ f(t)\in L^2(0,T) $ depends continuously on $ \{(\xi_n,\eta_n)\} $.

We introduce $ \zg_n=\xi_n+i\eta_n $ and
\begin{equation}
\ZLA{eq:DefidiZnMAIUSCOLO} \zn(t)= z_n(t)+ \intt H(t-s)z_n(s)\ZD s+in\intt K(t-s) z_n(s)\ZD s 
 \end{equation}
 so that the moment problem~(\ref{eq:MOMENTproblemDASTUDIARE}) takes the form
 \begin{equation}
\ZLA{eq:MOMENTproblemDASTUDIAREcompleNOSIMME}
\intT \zn(s) f(T-s)\ZD s=\zg_n\,,\qquad n>0\,.
\end{equation}

So,   the moment problem~(\ref{eq:MOMENTproblemDASTUDIARE}) is solvable and the solution $ f(t) $ depends continuously on the $ l^2 $ sequences $ \{\xi_n\} $ and $ \{\eta_n\} $ if and only if the sequence $ \{\zn(t)\} $ is a Riesz sequence in $ L^2(0,T) $. This we are going to prove now, and we shall see that any $ T\geq 2\pi $ will do.

 \subsection{Usefull estimates} 
The sequence $ \{z_n(t)\} $ has been studied in previous papers, in particular in  ~\cite{PandIEOT,PandDCDS1,PandDCDS2}, where we proved the following representation formula. Computing a second derivative of both the sides of~(\ref{eq:diZnPRIMORD}) we see that

\begin{eqnarray}\ZLA{eq:DiffeINTEseconrineBIS}
&&  z_n''(t)=2\zaa z_n'(t)-n^2 z_n(t)-n^2 \intt N'_{\zaa}(t-s) z_n(s)\ZD s\\
\nonumber &&\mbox{i.e. (using $ N'_{\zaa}(0)=0 $, see~(\ref{eq:lacondizioneNprimoZERO}))}\\
\ZLA{eq:DiffeINTEseconrineBIS}&&  z_n''(t)=2\zaa z_n'(t)-n^2N_{\zaa} (t)-n^2\intt N_{\zaa} (t-s)z_n'(s)\ZD s
\end{eqnarray}
and so  
\begin{align}\ZLA{eqINTdiz_n}
\nonumber z_n(t)= g_n(t)-\mu_n\left \{ \intt N'_{\zaa}(t-r) z_n(r)\ZD r \right.\\
\left.-\intt e^{\zaa s}\cos\zben s\left [
\int_0^{t-s}N_0(t-s-r)z_n(r)\ZD r
\right ]\ZD s\right \}
\end{align}
where
\begin{eqnarray*}
&& N_0(t)=N''_\zaa(t)-\zaa N'_\zaa(t)\,,\qquad g_n(t)=e^{\zaa t}\left [ \cos\zben t+\frac{\zaa}{\zben}\sin\zben t\right ]\,,\\
&& \zben =\sqrt{n^2-\zaa^2}\,,\qquad \mu_n=\frac{n^2}{\zben^2}\,.
\end{eqnarray*}
This equality holds with the possible exception of one index $ n_0 $: the exceptional index exists if    there exists a natural number $ n_0 $ such that $ \zaa^2=n_0^2 $, i.e.  $ \beta_{n_0} =0$.
In this case we have to replace the previous representation formula with the expression in~\cite[formula~(18)]{PandDCDS1}. We don't insist on this rather exceptional case here and we assume $ \zben\neq 0 $ for every $ n $. 

 We integrate by parts the last integral and we rewrite formula~(\ref{eqINTdiz_n}) as follows:
      \begin{eqnarray*}
    \nonumber  &&z_n(t)+\intt N_\zaa'(t-s) z_n(s)\ZD s=e^{\zaa t}\cos\zben t\\
 \ZLA{eq:PrimodaFAREmN}&& +\frac{\zaa}{\zben}\sin\zben t  +(1-\mu_n)\intt N_\zaa'(t-s) z_n(s)\ZD s\\
  \ZLA{eq:secondodaFAREmN}&& +\frac{1}{\zben}N_0(0)\mu_n \intt e^{\zaa(t-r)}\sin\zben (t-r) z_n(r)\ZD r\\
   \ZLA{eq:TERZOdaFAREmN}&&
  -\frac{1}{\zben}\mu_n \intt \left [\int_0^{t-r}e^{\zaa s}N_1(t-r-s)\sin\zben s\ZD s\right ] z_n(r)\ZD r.
      \end{eqnarray*}
       
      Here,
      \[
      N_1(t)=\zaa N_0(t)-N_0'(t)\,.
      \]
      
  We introduce $ L(t) $, the resolvent kernel of $ -N'_\zaa(t) $, given by
      \[ 
      L(t)=-\intt N'_\zaa(t-s)L(s)\ZD s-N'_\zaa(t)\,.
       \]
       We note that $ L(t) $ has the same regularity as $ N'_\zaa(t) $ and $ L(0)=0 $. Then we have the following equality:
       
       \begin{equation}\ZLA{eq:FormCONrisolv}
     z_n(t)= G_n(t)+\intt L(t-s)G_n(s)\ZD s
\end{equation}  where
  \begin{eqnarray}
    \nonumber  &&G_n(t)=e^{\zaa t}\cos\zben t\\
 \ZLA{eq:PrimodaFAREmN}&& +\frac{\zaa}{\zben}
 {  e^{\zaa t}}
 \sin\zben t  +(1-\mu_n)\intt N_\zaa'(t-s) z_n(s)\ZD s\\
  \ZLA{eq:secondodaFAREmN}&& + N_0(0) \frac{\mu_n}{\zben} \intt e^{\zaa(t-r)}\sin\zben (t-r) z_n(r)\ZD r\\
   \ZLA{eq:TERZOdaFAREmN}&&
  -\frac{\mu_n}{\zben} \intt \left [\int_0^{t-r}e^{\zaa s}N_1(t-r-s)\sin\zben s\ZD s\right ] z_n(r)\ZD r.
      \end{eqnarray}

We shall use the following result from~\cite[formulas~(2.14) and~(2.27)]{AvdoninPANDOLFI1}:
\begin{Lemma} For every $ T>0 $ there exists a number $ M $ such that for every $ n  $ we have:
 \begin{equation}
 \ZLA{eq:fattadomenica1}    
   |z_n(t)-e^{\zaa t}\cos\zben t|\leq \frac{M}{n}\,, \qquad 
    \left |\frac{z_n'(t)}{\zben}+e^{\zaa t}\sin \zben t \right |\leq \frac{M}{n} 
 \end{equation}
 (we can replace $ \zben $ with $ n $ in the previous formulas,  since $ \zben\asymp n $).
 In particular, the sequence $ \{z_n(t)\} $ is bounded on bounded intervals.
\end{Lemma}

Furthermore, using the representation~(\ref{eq:FormCONrisolv}) and 
\begin{equation}\ZLA{eq:diseQPerMUn}
1-\mu_n=\frac{\zaa^2}{n^2-\zaa^2}\,,\qquad \frac{n}{\zben}-1=\frac{\zaa^2}{\zben(n+\zben)} \,,
\end{equation}
  we see:
\begin{Lemma}\ZLA{Lemma:diseqPERintreg}
For every $ T $ there exists $ M_T $ such that
\begin{equation}
\ZLA{eq:diseqPERintreg}
\left |n \intt F(t-s) z_n(s)\ZD s-F(0)e^{\zaa t} \sin\zben t\right |\leq \frac{M}{n}
\end{equation}
for every     function $ F  \in H^2(0,T)$. If $ F\in H^1(0,T) $ then we have
\[ 
\sum \left |\left| n \intt F(t-s) z_n(s)\ZD s-F(0)e^{\zaa t} \sin\zben t\right |\right |^2_{L^2(0,T)}<+\ZIN\,.
 \]
\end{Lemma}
\zProof In this proof, $ \{M_n(t)\} $ denotes a sequence of functions which is bounded on $ [0,T] $ (not the same functions at every occurrence).

We use equality~(\ref{eq:FormCONrisolv}) and boundedness on $ [0,T] $ of the sequence $ \{z_n(t)\} $ to see that
\begin{eqnarray*}
&&n\intt F(t-\zt) z_n(\zt)\ZD\zt=n\intt F(t-\zt)e^{\zaa\zt}\cos\zben\zt\ZD\zt\\
&&+n\intt F(t-\zt)\int_0^\zt L(\zt-s)e^{\zaa s}\cos\zben s\ZD s\,\ZD \zt+\frac{M_n(t)}{\zben}\,.
\end{eqnarray*}
Using $ L(0)=0 $ and differentiability of $ F(t) $ and $ L(t) $, two 
  integrations by parts in the last integral shows that it can be absorbed  in $ M_n(t)/\zben $.
 
 We integrate by parts the first integral in the right hand side and we use $ (n/\zben)-1\asymp1/n^2 $ to see that
 \begin{eqnarray*}
 &&n\intt F(t-\zt)e^{\zaa\zt}\cos\zben\zt=F(0) e^{\zaa t}\sin\zben t\\
 &&+\frac{n}{\zben}\intt F'(t-\zt)e^{\zaa\zt}\sin\zben \zt\ZD \zt+ \frac{M_n(t)}{\zben}\,.
  \end{eqnarray*}
If $ F''(t)\in L^2 $ a further integration by parts shows that the last line is $ M_n(t)/\zben $. Otherwise we note that $\{ e^{i\zben t} \}$
 is a Riesz sequence in $ L^2(0,T) $ for every $ T\geq 2\pi $ (see~\cite[Appendix~5.1]{AvdoninPANDOLFI1}). Hence, the sequence $ \{\sin\zben t\} $ is Riesz on every interval $ L^2(0,T) $, $ T\geq \pi $ (the proof is similar to the corresponding proof for the cosine sequence given in~\cite{grubeev}).

We fix  $ T_0=\max\{\pi,T\} $ and we note that for every fixed $ t\in [0,T] $ we have
\[ 
\intt F'(t-\zt)e^{\zaa\zt}\sin\zben \zt\ZD \zt=\int_0^{T_0} \left [H(t-\zt)F'(t-\zt)e^{\zaa\zt}\right ]\sin\zben \zt\ZD \zt
 \]
where $ H(t) $ denotes the Heavisede function. Hence, for every fixed $ t $, these integrals are the ``Fourier'' coefficients of $  \left [H(t-\zt)F'(t-\zt)e^{\zaa\zt}\right ] $ in the expansion in terms of the biorthogonal of $ \{\sin\zben t\} $ and this gives (for a suitable constant $ M $)
\begin{eqnarray*}
&&
\sum _{n=1}^{+\ZIN} \left [\int_0^t \left [ F'(t-\zt)e^{\zaa\zt}\right ]\sin\zben \zt\ZD \zt\right ]^2\\
%%%%
&&=\sum _{n=1}^{+\ZIN} \left [\int_0^{T_0} \left [H(t-\zt)F'(t-\zt)e^{\zaa\zt}\right ]\sin\zben \zt\ZD \zt\right ]^2\\
&&\leq M\int_0^{T_0}\left [H(t-\zt)F'(t-\zt)e^{\zaa\zt}\right ]^2\ZD \zt 
 =M\intt\left | F'(t-\zt)e^{\zaa\zt}\right |^2\ZD \zt \,.
 \end{eqnarray*} 
A further integration from $ 0 $ to $ T $ gives the result.\zdia

\subsection{The proof of Theorem~\ref{teo:controllSTRESS/velocita}}

Statements~1 and~2 of Theorem~\ref{teo:controllSTRESS/velocita} are in Lemma~\ref{eq:lemmaCHEproVA1-2Teo1}. In order to prove the statement~3 we must prove that the sequence $ \{\zn(t)\} $ in~(\ref{eq:DefidiZnMAIUSCOLO}) is a Riesz sequence in $ L^2(0,T) $, provided that $ T\geq 2\pi $. This is the bulk of the proof, which requires several steps.

 It is convenient to introduce the following notations: $  {\mathbb Z}'={\mathbb Z}-\{0\} $ and, for $ n<0 $:
 \[ 
 \beta _{-n}=\zben\,,\qquad   \zg_n= \overline{\zg _{-n}}\,.
  \]
  Here $ \{\zg_n\}\in  l^2_{\mathbb C}({\mathbb Z}')$ (we shall denote $ l^2_{\mathbb C}({\mathbb Z}') $ simply as $ l^2 $).
  
So, both $ z_n(t) $ and $ \zn(t) $ are defined also for $ n<0 $ and

 \[ 
z_{n}(t)=z_{-n}(t)\,,\qquad   \zn(t)=\overline{Z _{-n}}(t)
  \]
  (since the memory kernels are real) and the moment problem~(\ref{eq:MOMENTproblemDASTUDIARE}) is equivalent to
  \begin{equation}
\ZLA{eq:MOMENTproblemDASTUDIAREcomplesso}
\intT \zn(t)f(T-t)\ZD t=\zg_n\,,\qquad n\in {\mathbb Z}'\,,\qquad \{\zg_n\}\in l^2\,.
\end{equation}
We are going to prove that $ \{\zn(t)\} _{n\in {\mathbb Z}'} $, is a Riesz sequence in $ L^2(0,T) $, $ T\geq 2\pi $.

The value of $ T\geq 2\pi $ is now fixed so that,
 using~(\ref{eq:DefidiZnMAIUSCOLO}) and Lemma~\ref{Lemma:diseqPERintreg} with $ F(t)=K(t) $ (and using $ K(0)=1 $)   we get:
\begin{equation}
\ZLA{eq:quadraticallyCLOSENESS}
  \zsumNoz \|\zn(t)- e^{(\zaa+i\zben)t}\|^2 _{L^2(0,T)}<+\ZIN\,.
\end{equation} 
As we noted, 
the sequence $ \{e^{(\zaa+i\zben)t}\} $ is a Riesz sequence in $ L^2(0,T) $ when $ T\geq 2\pi $ so that condition~(\ref{eq:quadraticallyCLOSENESS})  implies the existence of $ N $ such that $ \{\zn(t)\}_{|n|>N} $ is a Riesz sequence and so, using Bari Theorem   combined with~(\ref{eq:quadraticallyCLOSENESS}),     $ \{\zn(t)\}_{n\in{\mathbb Z}'} $ is a Riesz sequence in $ L^2(0,T) $ if and only if it is $ \omega $-independent, i.e. if and only if
    \[ 
    \zsumNoz \zaa_n\zn(t)=0 \ \implies\ \{\zaa_n\}=0\,.
     \]
  We recall that that the series here has to converge in $ L^2(0,T) $ and this is the case if and only if $ \{\zaa_n\}\in l^2 $. 
  
 We proceed in several steps to prove that $\{ \zn(t)\} $ is $ \ZOM $-independent.
  
  \subparagraph{Step 1: an equation for $ \zn(t) $.}

 Using~(\ref{eq:defiHK}) we see that
\begin{equation}\ZLA{eq:succeDAfareRIESZ}
   \zn(t)=
  z_n(t)+ in \intt N_\zaa(t-s)z_n(s)\ZD s+ \intt B_n(t-s) z_n(s)\ZD s 
   \end{equation}
   where
   \begin{equation}
   \ZLA{eq:definDiBn} B_n(t)=N_\zaa'(t)-2\zaa N_\zaa(t)+in\intt N_\zaa(t-r)M_\zaa(r)\ZD r\,.
\end{equation}
 Hence, using~(\ref{eq:diZnPRIMORD}),
 \[ 
 \zn(t)=\Lambda_nz_n(t)-\frac{i}{n}z_n'(t)+\intt B_n(s) z_n(t-s)\ZD s\,,\qquad \Lambda_n=(1+2\zaa i/n)\,.
  \]  
 We compute the derivative of both the sides, using~(\ref{eq:DiffeINTEseconrineBIS}). We get
 \begin{eqnarray*}
 &&\zn'(t)=\Lambda_n\left (2\zaa z_n(t)-n^2\intt N_\zaa(t-s) z_n(s)\ZD s\right )\\
 &&-\frac{i}{n}\left (2\zaa z_n'(t)-n^2N_\zaa(t)-n^2\intt N_\zaa(t-s)z_n'(s)\ZD s\right )\\
 &&+B_n(t)+\intt B_n (t-s)\left (
 2\zaa z_n( s)-n^2\int_0^{ s} N_\zaa( s-r)z_n(r)\ZD r
 \right )\ZD s\,.
 \end{eqnarray*}
 Collecting corresponding terms, we see that $ Z_n(t) $
 solves the following integrodifferential equation:
 \begin{eqnarray}
\nonumber&& \zn'(t)=2\zaa \zn(t)-n^2\intt N_{\zaa}(t-s)\zn(s)\ZD s+in N_\zaa(t)+B_n(t)\\
 \ZLA{eqperZn}
&&=2\zaa \zn(t)-n^2\intt N_{\zaa}(t-s)\zn(s)\ZD s+H(t)+in K(t)
\end{eqnarray}
 (the definitions of $ H(t) $ and $K(t)$ are in~(\ref{eq:defiHK})) and
$
  \zn(0)= 1 
  $.
   Note that this is similar 
   to~\cite[formula~(2.22)]{AvdoninPANDOLFI1}.
   
  \subparagraph{Step 2: The sequence $ \{\zn(t)\} $ is linearly independent in $ L^2(0,T) $ for every $ T>0 $.}
  The proof is by contradiction. If it is linearly dependent then there exist  $ N>0 $ and, corresponding to it,  an   index $ -K<0 $, and  coefficients $ \zaa_n\in {\mathbb C} $ such that
  \begin{equation}
\ZLA{eq:SelinearmDIPEND}
\sum _{n=-K}^N \zaa_n\zn(t)=0\qquad \mbox{in particular} \qquad \sum _{n=-K}^N \zaa_n =0    \,.
\end{equation}
\begin{Remark}{\rm
We recall that the indices are from $ {\mathbb Z}'$; i.e. $ n=0 $ is excluded. We can also include  $ n=0 $ in the sums, but then $ \zaa_0=0 $.\zdia}
\end{Remark}
We choose $ N >0$ to be the first index which corresponds to the minimum value of the numbers $ K> 0$.
Then we have also
\begin{multline}
0=\sum _{n=-K}^N \zaa_n\zn'(t) 
=2\zaa \sum _{n=-K}^N \zaa_n\zn(t)\\
-\intt N_{\zaa}(t-s)\left [\sum _{n=-K}^N n^2 \zaa_n \zn (s)\right ]\ZD s\\
\ZLA{eq:perLEDueCondSUzaa}+ H(t) \sum _{n=-K}^N  \zaa_n +iK(t) \sum _{n=-K}^N n \zaa_n  \,.
\end{multline}

The first series in the right hand side is zero (use the first equality in~(\ref{eq:SelinearmDIPEND})).  

Computing with  $ t=0 $ and using the second equality in~(\ref{eq:SelinearmDIPEND}), we get
\[
 \sum _{n=-K}^N n \zaa_n=0
 \]
so that the last line in~(\ref{eq:perLEDueCondSUzaa}) is zero i.e.  
   we have 
   
  \[ 
   \intt N_\zaa(t-s)\left [\sum _{n=-K}^N n^2 \zaa_n \zn (s)\right ]\ZD s=0\,.
   \]
Using that $ N_\zaa(t) $ is differentiable with $ N_\zaa(0)=1 $, we see that
 \[ 
 \sum _{n=-K}^N n^2 \zaa_n \zn (t)=0\,.
  \]
  This can be combined with~(\ref{eq:SelinearmDIPEND}) to see that an equality of the form~(\ref{eq:SelinearmDIPEND}) holds for $ N $ replaced by $ N-1 $, without increasing $ K $. 
  In fact we get
  \[ 
  \sum _{n=-K}^{N-1}(N-n)\zaa_n Z_n(t)=0\,.
   \]

  This contradicts the definition of $ N $.

 \subparagraph{Step 3: the sequence $ \{\zn(t)\} $ $ \ZOM $-independent, hence it is Riesz, in $ L^2(0,T) $ if $ T\geq 2\pi $} 
 
 We need the following Lemma, whose proof is in   Appendix~\ref{append:dimo:Lemma:diseqFOND}. 
 
 \begin{Lemma}\ZLA{Lemma:STIMAfondZn}
There exists a sequence $ \{M_n(t)\}_{n\in {\mathbb Z}'} $ of $ H^2 $ functions, for which the following properties hold:
\begin{itemize}
\item we have
\begin{equation}\ZLA{eq:NelLemma:STIMAfondZn}
Z_n(t)= e^{(\zaa +i\zben)t}+ M_n(t)\,.
\end{equation}
\item  the following series converge in $ L^2(0,T) $ for every sequence $ \{\zaa_n\}_{n\in {\mathbb Z}'}\in l^2 $ and for every $ T>0 $:  
\[ {\bf 1)}\  \zsumNoz \zaa_n M_n(t)\,,\qquad 
{\bf 2)}\ \zsumNoz  \zaa_n M'_n(t)\,,\qquad {\bf 3)}\  \zsumNoz \frac{\zaa_n}{\zben} M_n''(t)\,.
\]
  
\end{itemize}
\end{Lemma}

We recall that, in order to prove that $ \{\zn(t)\} $ is a Riesz sequence in $ L^2(0,T) $, we must prove that it is $ \ZOM $-independent. We proceed as follows: we assume that
a sequence $ \{\zaa_n\} $ satisfies
\begin{equation}
\ZLA{eq:UguaglSERIEaZERO}
\zsumNoz \zaa_n\zn(t)=0 
\end{equation}
in $ L^2(0,T) $ (so that necessarily $ \{\zaa_n\}\in l^2 $) and we prove $ \{\zaa_n\}=0 $. Relaying on Lemma~\ref{Lemma:STIMAfondZn}, we first prove the following additional ``regularity'' of the sequence $ \{\zaa_n\} $.
 
\begin{Lemma}\ZLA{Lemma:RegolaALPHAn}
Let~(\ref{eq:UguaglSERIEaZERO}) hold.   
 Then, there exists $ \{\zg_n\}\in l^2 $ such that
 \begin{equation}
\ZLA{eq:RegolALPHAn} 
\zaa_n=\frac{\zg_n}{\zben^2}\,.
\end{equation}
\end{Lemma}
\zProof 
In this proof we use the fact that $ \{e^{i\zben t}\}_{n\in {\mathbb Z}'} $ is a Riesz sequence in $ L^2(0,T) $ for every $ T\geq 2\pi $ of deficiency $ 1 $, and we get a Riesz basis if we add $ \beta_0=0 $ and we consider $ \{e^{i\zben t}\}_{n\in {\mathbb Z}} $, 
see the appendices in the papers~\cite{AvdoninPANDOLFI1,PandIEOT}. Consequently, from~\cite[Theorem~1]{Russel},
if we add a further exponential $ e^{ict} $ with $ c\neq \zben $ for every $ n\in {\mathbb Z} $ we get a Riesz basis of $ H^1(0,T) $, whose elements are the functions $ 1 $, $ e^{ct} $ and $ (1/\zben)e^{i\zben t} $ (here $ n\neq 0 $).

As we noted, convergence of the series in~(\ref{eq:UguaglSERIEaZERO}) implies that $ \{\zaa_n\}\in l^2 $. We multiply both the sides of~(\ref{eq:UguaglSERIEaZERO}) with $ e^{-\zaa t} $ and we use the representation~(\ref{eq:NelLemma:STIMAfondZn}) we see that
\begin{equation}\ZLA{eq:LemmaREGOLalphaN-1}
-\zsumNoz \zaa_ne^{i\zben t}=e^{-\zaa t}\zsumNoz \zaa_n M_n(t)\,.
 \end{equation}
Both the series converge in $ L^2(0,T) $ for every $ T $ and Lemma~\ref{Lemma:STIMAfondZn} asserts that the series  $ \zsumNoz \zaa_n M_n'(t)  $ converges in $ L^2(0,T) $ too. Hence, it represents an $ H^1(0,T) $ functions, which can be expanded in series of $ 1 $, $ e^{ict} $ and $ (1/\zben) e^{i\zben t} $. So we have also
\[ 
-\zsumNoz \zaa_ne^{i\zben t}=e^{-\zaa t}\zsumNoz \zaa_n M_n(t)=\ZDE_0
+ \ZDE_c e^{ict}
+\zsumNoz \ZDE_n \frac{1}{\zben} e^{i\zben t}\,.
 \]
Equating the corresponding coefficients we see that $ \ZDE_0=0 $, $ \ZDE_c=0 $ and 
\[ 
\zaa_n=\frac{\ZDE_n}{\zben} 
 \]
so that
\begin{equation}
\ZLA{eq:LemmaREGOLalphaN-2}
-\zsumNoz \frac{\ZDE_n}{\zben}e^{i\zben t}=e^{-\zaa t}\zsumNoz \frac{\ZDE_n}{\zben} M_n(t)\,,\qquad \{\ZDE_n\}\in l^2\,.
\end{equation}
Now we compute the derivatives of both the sides of~(\ref{eq:LemmaREGOLalphaN-2}) and we get:
\begin{equation}
\ZLA{eq:LemmaREGOLalphaN-3}
-i\left [\zsumNoz  \ZDE_n  e^{i\zben t}\right ]=e^{-\zaa t}\zsumNoz \frac{\ZDE_n}{\zben} M_n'(t)-\zaa e^{-\zaa t}\zsumNoz \frac{\ZDE_n}{\zben} M_n(t)
\,,\qquad \{\ZDE_n\}\in l^2\,.
\end{equation}

Our assumption now is that  the series
\[ 
\zsumNoz \frac{\ZDE_n}{\zben} M_n'(t)\,,\qquad \zsumNoz \frac{\ZDE_n}{\zben} M_n''(t)
 \]
 converge  in $ L^2(0,T) $ and so the right hand   side of~(\ref{eq:LemmaREGOLalphaN-3})  belongs to $ H^1(0,T) $. As above, being $ T\geq 2\pi $, we have
 \[ 
 \ZDE_n=\frac{\zg_n}{\zben}\,,\quad {\rm i.e.}\quad  \zaa_n=\frac{\zg_n}{\zben^2}\,,\qquad \{\zg_n\}\in l^2\,.\zdiaform
  \]

We use this lemma as follows. Equality~(\ref{eq:RegolALPHAn}) implies convergence  of the following series, which then have to converge to $ 0 $:
\[
\zsumNoz \frac{\zg_n}{\zben^2} \zn(t)=0\,,\qquad \zsumNoz \frac{\zg_n}{\zben^2} =0\,, \qquad \frac{\ZD}{\ZD t}\left [
\zsumNoz \frac{\zg_n}{\zben^2}  \zn(t)
\right ]=\zsumNoz \zaa_n \zn'(t) =0\,.
 \]
So, using~(\ref{eqperZn}) we get
\begin{eqnarray*}
%\ZLA{eq:IntermeSERIE}
&&  \intt N(t-s)\left [ \zsumNoz \zg_n \zn(s) \right ]\ZD s =H(t) \zsumNoz  \frac{\zg_n}{\zben^2}\\
 &&+iK(t)\zsumNoz \frac{n}{\zben} \frac{\zg_n}{\zben}
\end{eqnarray*}
As we noted, the first series on the right side vanishes. 
Equality~(\ref{eq:RegolALPHAn}) shows convergence of the last series and also it shows that we can compute both the sides with $ t=0 $. We get
\[ 
  \zsumNoz \frac{n}{\zben} \frac{\zg_n}{\zben}=0
 \]

 So we have
 \begin{equation}\ZLA{eq:seNONomegINDIPderiv}
  \intt N(t-s)\left [ \zsumNoz  
 \zg_n \zn(s) \right ]\ZD s   =0 \quad {\rm i.e.}\quad 
   \zsumNoz  
 \zg_n \zn(t) =0\,.
  \end{equation}
  We can combine~(\ref{eq:UguaglSERIEaZERO}) and~(\ref{eq:seNONomegINDIPderiv}) so to get a new series
  \[ 
  \sum _{n\neq 0\,,\ n\neq 1}\zaa^{(1)}_n \zn(t)=0\qquad \left \{\zaa^{(1)}_n\right \}=\left \{
  \zg_n\left (
  \frac{1}{\beta_1^2}-\frac{1}{\zben^2}
  \right )
  \right \}\in l^2\,.
   \]
   
   Note that if $ n\neq 1 $ then $ \zaa_n^{(1)}=0 $  if and only if $ \zaa_n=0 $.
   We repeate this procedure till we remove $ N -1$ positive and $ N-1 $ negative coefficients and we end up with the equality
   \[ 
    \sum _{|n|\geq N}\tilde\zaa _n \zn(t)=0
    \]
  and so $ \{\tilde \zaa_n\}=0 $ since $ \{ \zn(t)\} _{|n|\geq N} $ is a Riesz sequence. As $ \tilde \zaa_n=0 $ if and only if $ \zaa_n=0 $, we see that the series~(\ref{eq:UguaglSERIEaZERO}) is a finite sum, and so we have also $ \zaa_n=0 $ if $ |n|<N $ since the sequence $ \{\zn(t)\} $ is linearly independent.

    \begin{Remark}
     {\rm
 Finally we note that    Lemma~\ref{Lemma:STIMAfondZn} can be interpreted as a weaker version of the asymptotic estimates in~\cite[Lemma~5.3]{AvdoninPANDOLFI1}, which required one more derivative of $ N_\zaa(t) $.\zdia
     }
     \end{Remark}

\section{\ZLA{sect:stress-deformation}Deformation and stress}

Now we examine the pair of the deformation $ w(\cdot;T) $ and the stress $ \ZSI(\cdot;T) $. Let us go back to the 
series  of the deformation and the series of the stress in~(\ref{serieDIflussoVELOCITA})  which converge respectively in $ L^2(0,\pi) $ and $ H^{-1}(0,\pi) $. We shall see that the sequence of the  Fourier  coefficients of this series are asymptotic one to the other. Namely we shall prove:
\begin{Theorem}\ZLA{teo:stress-deformation}
  Let $ T>0 $ and let  $ f(t) $ be a control which drives the deformation from the initial condition $ w(\cdot,0)=0 $ to $ w(\cdot, T) $. Let $ w_n $ be the  Fourier  coefficients of $ w(\cdot, T) $, i.e. 
\begin{equation}
\ZLA{eq:secDefoSTReeq1}
w_n= \intT f(T-r)\left \{ n \intr N_\zaa(r-s) z_n(s)\ZD s\right \}\ZD r \,.
\end{equation}

 Let
 \begin{equation}
\ZLA{eq:secDefoSTReeq2}
 \ZSI_n=\intT f(T-r)\left \{ n\intr K(r-s)z_n(s)\ZD s\right \}\ZD r
\end{equation}
  ($ K(t) $ defined in~(\ref{eq:defiHK})).

Then,   there exists a number $ M $ (which depends on $ T $ and $ f $)  such that
  \[ 
  |\ZSI_n-w_n|\leq\frac{M}{n}\,.
   \]
\end{Theorem}
\zProof In this theorem we are not assuming that $ T $ has to be ``large''. We assume only that it is positive. So, even the sole sequence $ \{w_n\} $ in~(\ref{eq:secDefoSTReeq1}) will not be arbitrary. Furthermore, the sequencese $ \{w_m\} $ and $ \{\ZSI_n\} $ do not uniquely identify the function $ f $. So, let us consider   one special $ f\in L^2(0,T) $ for which equalities~(\ref{eq:secDefoSTReeq1}) and~(\ref{eq:secDefoSTReeq2})  both hold.

We observe that
\begin{eqnarray*}
&&\ZSI_n=\intT f(T-r)\left \{ n\intr N_\zaa(r-s) z_n(s)\ZD s\right \} \ZD r\\
&&+ \intT f(T-r)\left \{
n\intr z_n(s)\int_0^{r-s}N_\zaa(r-s-\nu) M_\zaa(\nu)\ZD \nu\,\ZD s
\right \}\ZD r\\
&&= w_n+ \intT f(T-r)\left \{
n\intr z_n(s)\int_0^{r-s}N_\zaa(r-s-\nu) M_\zaa(\nu)\ZD \nu\,\ZD s
\right \}\ZD r\,.
\end{eqnarray*}
The last integral is less then
\begin{equation}\ZLA{eq:DiniTioNofL}
\| f\|_{L^2(0,T)}\left \|
n\intr F(r-s) z_n(s) \ZD s
\right \|_{L^2(0,T)}\,,\quad F(t)=\intr N_\zaa(t -\nu) M_\zaa(\nu)\ZD \nu\,.
 \end{equation}
The function $ F(t) $ is twice differentiable and moreover $ F(0)=0 $
 so that, using Lemma~\ref{Lemma:diseqPERintreg},
 
 \[ 
 \left |\left | n\intt F(t-s) z_n(s)\ZD s\right |\right |_{L^2(0,T)}\leq \frac{M}{n}\,.
  \]
  The result follows from here.\zdia
  
 Up to now the results we have found parallel those of the purely elastic case. Now we can observe a difference, which might have some interest, concerning the long wavelength  components. Clearly components which correspond to $ \phi_n(x) $ with ``large'' $ n $, i.e. short wavelength, as computed by model~(\ref{eq:equaSECord1}) will not represent the real behavior of the system, due to unmodeled dynamics, dissipations etc., not taken into account when deriving Eq.~(\ref{eq:equaSECord1}) and only the first components will be (hopefully) realistic. And here we have a difference with the purely elastic case, since in the purely elastic case the generalized  Fourier  coefficients cannot be assigned at will for the deformation and stress, not even for a single wavelength. Instead:
 \begin{Theorem}
Let $ N>0 $ be fixed and let $ \{c_n\} $ and $ \{d_n\} $ be two finite sequences of real numbers, $ 1\leq n\leq N $. Then there exists a function $ f(t)\in L^2(0,T) $ which assign the ``Fourier'' coefficients $ \{c_n\} $ to the deformation and $ \{d_n\} $ to the stress.
\end{Theorem}
\zProof In order to prove the theorem, it is sufficient that we show solvability of the following (finite) moment problem (the function $ F(t) $ is defined in~(\ref{eq:DiniTioNofL})):
\[ 
\intT f(T-r)\left [ \intr \left (N_\zaa(r-s)+i F(r-s)\right ) z_n(s)\ZD s\right ]\ZD r=\dfrac{1}{n}\left [c_n+i (d_n-c_n)\right ]\,.
 \]
 This problem is solvable if and only if the functions 
 \[ \intr \left (N_\zaa(r-s)+i F(r-s)\right ) z_n(s)\ZD s 
 \]
 are linearly independent.
 
 Assume not. Then there exist numbers $ \zaa_n $ such that
 \[ 
 \sum _{n=1}^N \zaa_n\left [
 \intr \left (N_\zaa(r-s)+i F(r-s)\right ) z_n(s)\ZD s 
 \right ]=0\,.
  \]
  We use $ N(0)=1 $ and $ F(0)=0 $ and we compute the derivatives of both sides, which is zero. hence we have
  \[ 
   \sum _{n=1}^N \zaa_n z_n(r)=-\intr \left (N_\zaa'(r-s)+i F'(r-s)\right )\left [ \sum _{n=1}^N \zaa_n z_n(s)\right ]\ZD s\,.
   \]
 Uniqueness of solution of Volterra integral equations implies
    \[ 
        \sum _{n=1}^N \zaa_n z_n(r)  =0
     \]
     and this is possible only when each coefficient $ \zaa_n=0 $, because the sequence $ \{z_n(t)\} $ is linearly independent, see~\cite{PandDCDS1}.\zdia

   \appendix

     \section{\ZLA{appendix:prooflemma:complettezzaH-1}Appendix: the proof of Lemma~\ref{lemma:complettezzaH-1}}

\zProof   
The first statement (which could also be proved as the second one) is easily seen because $ \{(\sqrt{2/\pi})\sin nx\} $
is the orthonormal basis of $ L^2(0,\pi) $, of the eigenvectors of the operator $ Au=\zthe _{xx} $ with domain $ H_0^1(0,\pi)\cap H^2(0,\pi) $ and $ H_0^1(0,\pi) $ is $ {\rm dom }\, (-A)^{1/2} $. Hence,   $ \{(1/n)(\sqrt{2/\pi})\sin nx \} $ is a orthonormal basis of $ H_0^1(0,\pi) $ while $\{n(\sqrt{2/\pi})\sin nx\}   $ is a orthonormal basis of its dual $ H^{-1}(0,\pi) $.

In order to prove the second statement we 
show that any $ \chi\in H^{-1}(0,\pi) $ can be represented as
\[ 
\chi =\sum _{n=1}^{+\ZIN} c_n \left (n\cos nx\right )
 \]
(so that the sequence $ \{n\cos nx\} $ is complete in $ H^{-1}(0,\pi) $) and that there exist $ m>0 $ and $ M $ such that
\begin{equation}\ZLA{eq:perRieszCOSenO}
 m \sum _{n=1}^{+\ZIN} |c_n|^2\leq \|\chi\|^2 _{H^{-1}(0,\pi)}\leq M \sum _{n=1}^{+\ZIN} |c_n|^2 
\end{equation} 
 (i.e., the proof relays on Theorem~\ref{Teo:BARIcarattRIESZ}).
 
 To be more precise   the distributions $ n\cos nx $  in this formula are distributions   on $ (-\pi,\pi) $, localized to $ (0,\pi) $.
 
 For clarity, we introduce the notations
 \[ 
 \ZZL\cdot,\cdot\ZZR _{(-\pi,\pi)}\,,\qquad  \ZZL\cdot,\cdot\ZZR_{(0,\pi)}
  \]
  to denote the pairings of respectively $ H^1_0(-\pi,\pi) $ and $ H^1_0(0,\pi) $ and their duals $ H^{-1}(-\pi,\pi) $ and $ H^{-1}(0,\pi) $.
  
  We represent $ \phi\in H^1_0(-\pi,\pi) $ as
  \[ 
  \phi(x)=\phi_p(x)+\phi_d(x)+f(x)\phi(0)\,,\qquad \left\{
  \begin{array}{l}
f(x)=1-(x/\pi)^2\\
\phi_d(x)=\frac{1}{2}\left [ \phi(x)-\phi(-x)\right ]\\
\phi_p(x)=\frac{1}{2}\left [\phi(x)+\phi(-x)\right ]-f(x)\phi(0)\,.
\end{array} \right.
   \] 
 Once the function $ f(x) $ has been fixed, this representation is unique and   the restrictions to $ [0,\pi] $ of the functions $ \phi_p(x) $ belongs to $ H^1_0(0,\pi) $. 
   
   We associate to each $ \chi\in H^{-1}(0,\pi) $ the distribution $ \chi_e  \in H^{-1}(-\pi,\pi) $ defined as follows:
   \[ 
   \ZZL \chi_e  ,\phi\ZZR _{(-\pi,\pi)}=\frac{1}{2}\left [ \ZZL \chi ,\phi(x)-\phi(0)f(x)\ZZR _{(0,\pi)}+ \ZZL \chi ,\phi(-x)-\phi(0)f(x)\ZZR _{(0,\pi)}\right ]
    \]
   With this definition,
 \begin{equation}
\ZLA{eq:DellaDISPARita}
   \ZZL \chi_e  ,\phi(x)\ZZR _{(-\pi,\pi)}=   \ZZL \chi_e  ,\phi(-x)\ZZR _{(-\pi,\pi)}\qquad \forall \phi\in H^1_0(-\pi,\pi)  
\end{equation}
i.e., any distribution $ \chi_e $ is an even element of $ H^{-1}(-\pi,\pi) $.

The transformation $ \chi\mapsto\chi_e   $ from $ H^{-1}(0,\pi) $ to $ H^{-1}(-\pi,\pi) $ is  (linear and) continuous. In fact we have
\[ 
\|\chi_e\|_{ H^{-1}(-\pi,\pi)}\leq 8\|\chi \|_{ H^{-1}(0,\pi)}\,.
 \]
 
 Let $ \phi(x)\in H^{1}_0(0,\pi) $ and $ \phi_e(x) $ its even extension to $ [-\pi,\pi] $.  Then $ \phi(0)=\phi_e(0)=0 $ and 
 \[ 
  \ZZL \chi_e  ,\phi_e(x)\ZZR _{(-\pi,\pi)}= \ZZL \chi,\phi(x)\ZZR _{(0,\pi)}\,.
  \]
   
 Instead, if we consider the extension $ \phi_l(x) $ with $\phi _l(x)=0 $ for $ x\leq 0 $ then we have
 \begin{equation}
\ZLA{eq:perRESTRI}
  \ZZL \chi_e  ,\phi_l(x)\ZZR _{(-\pi,\pi)}= \frac{1}{2}\ZZL \chi,\phi(x)\ZZR _{(0,\pi)} 
  \end{equation}
 i.e., $ \chi $ is twice the localization of $ \chi_e $ to $ (0,\pi) $,  applied to the elements of $ H^1_0(0,\pi) $.
 These equalities  in particular show that $ \chi $ can be reconstructed from $ \chi_e   $ (and so the transformation $ \chi\mapsto \chi_e $ is injective) as follows: in order to compute $ \ZZL \chi,\phi(x)\ZZR _{(0,\pi)} $
 first we extend $ \phi $ to $ \phi_l $. Note that
 \[ 
 \|\phi\|_{H^1_0(0,\pi)}=\|\phi_l\|_{H^1_0(-\pi,\pi)}\,.
  \]
  Then we compute $\ZZL \chi_e  ,\phi_l(x)\ZZR _{(-\pi,\pi)}  $. So,
  \begin{eqnarray} 
\nonumber && \|\chi\| _{H^{-1}(0,\pi)}=\sup _{\|\phi\|_{H^1_0(0,\pi)}=1}
     \ZZL \chi,\phi(x)\ZZR _{(0,\pi)}\\
 \ZLA{eq:relaNORMECHIeCHId}    &&=2\sup _{\|\phi_l\|_{H^1_0(-\pi,\pi)}=1}
  \ZZL \chi_e  ,\phi_l(x)\ZZR _{(-\pi,\pi)}\leq 2\|\chi_e  \|_{H^{-1}(-\pi,\pi)}
 \end{eqnarray}
 So, the transformation from $ \chi $ to $ \chi_e   $ is an isomorphism from $H^{-1}(0,\pi)$ to its image, contained in the subspace the ``even'' distribution in $ H^{-1}(-\pi,\pi) $: with   $ m_0=1/8$ and $ M_0=2 $ we have
 \begin{equation}\ZLA{eq:RelaNORMEconSENZAe}
 m_0\|\chi_e\| _{H^{-1}(-\pi,\pi)}\leq \|\chi \| _{H^{-1}(0,\pi)}\leq M_0 \|\chi_e\| _{H^{-1}(-\pi,\pi)}\,.
  \end{equation} 
  
    From~\cite{Russel}, we know that $ \left \{ ne^{inx}\right \}_{n\neq 0} $ is a Riesz basis of $ H^{-1}(-\pi,\pi ) $
so that we can write
\begin{equation}
\ZLA{eq:peRLARieszExpodaRussel}
\chi_e  =\sum _{n\neq 0} c_n \left (n e^{inx}\right )\,,\qquad  \tilde m\sum _{n\neq 0} |c_n|^2\leq \|\chi_e  \|^2_{H^{-1}(-\pi,\pi)}\leq \tilde M \sum _{n\neq 0} |c_n|^2
\end{equation}
 ($ \tilde m>0 $ and $ \tilde M $ suitable constants). 
 From~(\ref{eq:DellaDISPARita}) we have
  \begin{align*}
  \sum _{n\neq 0}c_n \ZZL \left ( ne^{in x}\right ),\phi(x)\ZZR _{(-\pi,\pi)}=
  \sum  _{n\neq 0}c_n \ZZL \left ( ne^{in x}\right ),\phi(-x)\ZZR _{(-\pi,\pi)}\\
  =  \sum _{n\neq 0}c_n \ZZL \left ( ne^{-in x}\right ),\phi(x)\ZZR _{(-\pi,\pi)}
  = 
  \sum _{n\neq 0}(-c_{-n}) \ZZL \left ( ne^{in x}\right ),\phi(x)\ZZR _{(-\pi,\pi)}
  \end{align*}
   and so
   \[ 
   c_n=-c _{-n}\,.
    \]
   This shows that

\[
   \chi_e  =\sum _{n=1}^{+\ZIN} (2c_n) \left( n\cos nx\right )\,.
    \]
   Now we combine~(\ref{eq:RelaNORMEconSENZAe})  and~(\ref{eq:peRLARieszExpodaRussel})  in order to get  
   \[ 
  \tilde mm_0^2\sum _{n\neq 0} |c_n|^2
  \leq m_0^2\| \chi_e\|^2_{H^{-1}(-\pi,\pi)}
  \leq \|\chi\|^2_{H^{-1}(0,\pi)}\leq M_0^2\|\chi_e\|^2_{H^{-1}(-\pi,\pi)}\leq  \tilde M M_0^2 \sum _{n\neq 0} |c_n|^2\,.
    \] 
    
   The result now follows since from~(\ref{eq:perRESTRI}), we have
   
   \[ 
   \ZZL\chi,\phi\ZZR_{(0,\pi)} =\left\langle\!\left \langle \sum _{n=1}^{+\ZIN} 4c_n{\cal L} (n\cos nx),\phi\right \rangle\!\right \rangle_{(0,\pi)}
    \]
  for every $ \phi\in H^1_0(0,\pi) $,  where  $ \cal L $ denotes the localization of a distribution on $ (-\pi,\pi) $ to the interval $ (0,\pi) $.

\begin{Remark}
{\rm The properties of the sine and cosine sequences, one an orthonormal basis and the second  a Riesz basis, can be interchanged, since every Riesz basis is an orthonormal basis  with a suitable, equivalent, Hilbert space norm.\zdia}
\end{Remark}

      \section{\ZLA{append:dimo:Lemma:diseqFOND}Appendix: the proof of Lemma~\ref{Lemma:STIMAfondZn}}

      We denote $ \{M_n(t)\} $ a sequence of functions with the properties {\bf 1)}, {\bf 2)}, {\bf 3)} stated in the second item of Lemma~\ref{Lemma:STIMAfondZn}. Unless needed for clarity, we don't distinguish among different occurrencies of these functions, so $ M_n(t) $ is not the same at every occurrence.

       \begin{Remark}\ZLA{Rema:simple-observation}
       {\rm Properties \textbf{1)}, \textbf{2)} and \textbf{3)} of a sequence $ \{M_n(t)\} $ are retained under convolution with a fixed integrable kernel.
\zdia}
\end{Remark}
      
    Using formula~(\ref{eq:succeDAfareRIESZ}), we represent $ \zn(t) $ as
    \[ 
    \zn(t)=z_n(t)+in\intt R_n(t-s) z_n(s)\ZD s\,,\qquad R_n(t)=K(t)-i \frac{1}{n}H (t)
     \]
     and we prove the existence of two sequences $ \{\tilde M_n(t)\} $, $ \{\hat M_n(t)\}$ such that
     \begin{equation}\ZLA{eq:AppeBleDUEformule} 
     z_n(t)=e^{\zaa t} \cos\zben t+\tilde M_n(t)\,,\qquad 
     n\intt R_n(t-s) z_n(s)\ZD s=e^{\zaa t} \sin\zben t+\hat M_n(t)\,.
      \end{equation}  Then,   $ M_n(t)=\tilde M_n(t)+\hat M_n(t) $.
      
      We prove the first equality in~(\ref{eq:AppeBleDUEformule}). We relay on~(\ref{eq:FormCONrisolv}). 
       First 
      we prove that every term in the rows~(\ref{eq:PrimodaFAREmN})-(\ref{eq:TERZOdaFAREmN}) has the properties of an $ M_n(t) $ so that, by applying the observation in Remark~\ref{Rema:simple-observation} to the convolution in~(\ref{eq:FormCONrisolv})
       we shall get
    {   \[ 
   \left |   z_n(t)-e^{\zaa t}\cos\zben t\right  |\leq M_n(t)\,.
       \]  }  %fine colore
       In fact, it is clear that
       \[ 
       \intt L(t-s) e^{\zaa s}\cos\zben s\ZD s=M_n(t)
        \]
        since $ L'(t)\in H^1(0,T) $.
 As to the   first function in~(\ref{eq:PrimodaFAREmN}), we clearly have $ \frac{\zaa}{\zben}
  {  e^{\zaa t}}
 \sin\zben t =M_n(t)$.     
 
  The second function in~(\ref{eq:PrimodaFAREmN}) has the first and second derivative given by
   \[
(1-\mu_n)    \intt N_\zaa ''(t-s) z_n(s)\ZD s\,,\qquad (1-\mu_n) \left \{ N_\zaa''(t)+\intt N_\zaa ''( s) z_n'(t-s)\ZD s\right \}
   \]
   so that the required properties \textbf{1)}, \textbf{2)} and \textbf{3)} are clear, 
   using the second inequality in~(\ref{eq:fattadomenica1}) and $ 1-\mu_n=\zaa/(n^2-\zaa^2) $.
      
      We consider row~(\ref{eq:secondodaFAREmN}). We study the   series
 \[ 
\zsumNoz \zaa_n
 \frac{\mu_n}{\zben}\intt e^{\zaa(t-r)} \sin\zben(t-r) z_n(r)\ZD r
  \]   
     which is uniformly convergent, since $ \{ z_n(t)\} $ is bounded on bounded intervals. 
     
     Computing the first derivatives we have two series:
     \begin{equation}
\ZLA{PerLamNdAppendI}
 \begin{array}{l}
\displaystyle  \zsumNoz \zaa_n
 \frac{\mu_n}{\zben}  \zaa\intt e^{\zaa(t-r)}\sin\zben(t-r) z_n(r)\ZD r
 \\[3pt]
\displaystyle \zsumNoz \zaa_n\mu_n\intt e^{\zaa(t-r)}\cos\zben (t-r) z_n(r)\ZD r\,.\end{array} 
\end{equation}
Uniform convergence of the first series is clear. The second series is elaborated using the first inequality in~(\ref{eq:fattadomenica1}) which we write as
\[ 
z_n(t)=e^{\zaa t}\cos \zben t+\frac{1}{n}H_n(t)
 \]
and $ \{ H_n(t)\} $ is bounded on $ [0,T] $. So,
\begin{eqnarray*}
&& \zsumNoz \zaa_n\mu_n\intt e^{\zaa(t-r)}\cos\zben (t-r) z_n(r)\ZD r\\
&&= e^{\zaa t}\zsumNoz \zaa_n\mu_n\left [ \frac{1}{2}t\cos\zben t+\frac{1}{2\zben}\sin\zben t\right ]\\
&&+\zsumNoz  \frac{ \zaa_n}{n}\mu_n\intt e^{\zaa(t-r)}\cos\zben(t-r) H_n(r)\ZD r\,.
\end{eqnarray*}
$ L^2(0,T) $-convergence is now clear.

So, properties {\bf 1)} and {\bf 2)} hold. In order to prove property {\bf 3)} replace $ \zaa_n $ with $ \zaa_n/\zben $ in the series~(\ref{PerLamNdAppendI}), and compute the second derivative. The series that we get can be treated as above.

 Finally, we consider the term~(\ref{eq:TERZOdaFAREmN}). Properties  {\bf 1)} and {\bf 2)} are obvious, because the derivative of the term at the line~(\ref{eq:TERZOdaFAREmN}) is the sum of the following terms:
 \begin{equation}
 \ZLA{eq:Appe-2TerTERM}\begin{array}{l}
 \displaystyle  \zaa\frac{\mu_n}{\zben}\intt \left [\int_0^{t-r}e^{\zaa(t-r-s)}N_1(s)\sin\zben(t-r-s)\ZD s\right ] z_n(r)\ZD r\\[3pt]
  \displaystyle  \mu_n\intt\left [ \int_0^{t-r} e^{\zaa(t-r-s)}  N_1(s)\cos\zben(t-r-s)\ZD s\right ] z_n(r)\ZD r\,. 
  \end{array}
  \end{equation}
  The first term has the property of an $ M_n(t) $ and the same holds for the second too, since 
  \[ 
  \intt N_1(t-s) e^{\zaa s}\cos\zben s\ZD s
   \]
   is the Fourier coefficient (in a cosine series) of $ N_1(t-s) e^{\zaa s} {\bf 1}(t-s) $ where $ {\bf 1}(t)  $ is the Heaviside function (see the argument in the proof of Lemma~\ref{Lemma:diseqPERintreg}). Hence,
   \[ 
  \zsumNoz   \zaa_n \int_0^T\left |
   \intt N_1(t-s)e^{\zaa s}\sin\zben s\ZD s
   \right |^2\ZD t\leq +\infty\,.
    \]
   This shows properties {\bf 1)} and {\bf 2)} of the second term in~(\ref{eq:Appe-2TerTERM}).   
  We note now property $ {\bf 3)} $. In fact, the second derivative is the sum of the following terms:  
 
  \begin{eqnarray*}
  &&
  \mu_n\intt N_1(s) z_n(t-s)\ZD s\,,\\
  && \zaa\mu_n \intt N_1(s)\int_0^{t-s}e^{\zaa(t-s-r)}\cos\zben(t-s-r)z_n(r)\ZD r\,\ZD s\,,\\
  && 
-\zben\mu_n  \intt N_1(s)\left [
  \int _0^{t-s} e^{\zaa(t-s-r) } \sin\zben(t-s-r)z_n(r)\ZD r
  \right ]\ZD s\,.
   \end{eqnarray*}
The corresponding series (with coefficients $ \zaa_n/\zben $) is seen to be $ L^2 $-convergent, using arguments similar to the previous ones.

      This gives the required property of the first addendum on the right hand side of~(\ref{eq:DefidiZnMAIUSCOLO}).  The fact to be noted is that in these computations the regularity used is $ M\in H^2 $.

      Now we consider  the convolution of $ z_n(t) $ with the functions
     $ H(t) $ and $      nK(t)$ 
       (we recall $ K(0)=1 $).
       
       The convolution of $ H(t) $ with every term in the right hand side of~(\ref{eq:FormCONrisolv})       
        gives a sum of terms which have the properties of $ M_n(t) $. This is
         clear, using Remark~\ref{Rema:simple-observation}, for every term, a part possibly the first one

         \[ 
         \intt H(t-s) e^{\zaa s}\cos\zben s\ZD s\,,
          \]
         
         whose first and second derivatives are respectively
       \begin{eqnarray*}
       &&H(0) e^{\zaa t } \cos\zben t+\intt H'(s) e^{\zaa(t-s)}\cos\zben (t-s)\ZD s\,,\\
       &&   e^{\zaa t}\biggl [\zaa H(0)\cos\zben t-\zben H(0)\sin\zben t+ \zaa \intt e^{\zaa(t- s)}H'(t-s) \cos\zben s \ZD s \biggr. \\
       &&
        \biggl.
        -\zben \intt e^{\zaa(t-s)}H'(t-s) \sin\zben(t-s)\ZD s\biggr]
       \end{eqnarray*}
       from which the properties of being $ M_n(t) $ are clear.

         Now we consider the convolution with $ nK(t) $.

         The first addendum gives
         \[ 
         n\intt K(t-s) e^{\zaa s}\cos\zben s\ZD s=e^{\zaa t}\sin\zben t +M_n(t)
          \]
          where 
          
          \[ 
          M_n(t)=\left (\frac{n}{\zben}-1\right )e^{\zaa t}\sin\zben t-\frac{n}{\zben } \intt e^{\zaa s}\left [\zaa K(t-s)-K'(t-s)\right] \sin\zben s \ZD s\,.
           \]
   The notation $ M_n(t) $ to denote this term is legitimate, as seen  using $ K\in H^{2} $ and $ (n/\zben-1)\asymp 1/n^2 $.   
      
      Finally, the convolution of $ nK(t) $  with the functions at the row~(\ref{eq:PrimodaFAREmN})-(\ref{eq:TERZOdaFAREmN}) contributes terms with the properties of $ \{M_n(t)\} $. The computations are similar to the previous ones, and are left to the reader. We confine ourselves to note that in these computations we encounter integrals of the form~(\ref{PerLamNdAppendI}).

\enddocument